\documentclass[twocolumn,showpacs,preprintnumbers,amsmath,amssymb,superscriptaddress]{revtex4-1}
\usepackage[latin1]{inputenc}
\usepackage{amsmath}
\usepackage{amssymb}
\usepackage{graphicx}  
\usepackage{dcolumn}   
\usepackage{bm}        
\usepackage{array}
\usepackage{hyperref}
\hypersetup{colorlinks=true,%
linkcolor=blue,%
pdfproducer={Birds-of-a-Feather},%
pdfstartview=FitB}

\begin{document}
\title{Level spacing distribution of a Lorentzian matrix at the spectrum edge }

\author{Adel Abbout}
\affiliation{Service de Physique de l'\'Etat Condens\'e (CNRS URA 2464),
IRAMIS/SPEC, CEA Saclay, 91191 Gif-sur-Yvette, France.\\
Laboratoire CRISMAT, UMR 6508 CNRS, ENSICAEN et Universit\'e de Caen Basse Normandie, 6 Boulevard Mar\'echal Juin,
F-14050 Caen, France.}
\author{Peng Mei}
\affiliation{University of Helsinki, Department of Mathematics and Statistics
P.O. Box 68, FI-00014 Helsingin yliopisto, Finland
}

\begin{abstract}

Effective Hamiltonians can explain in a much simpler way the physics behind a scattering process. Chaotic scattering is directly related to Lorentzian Hamiltonians  which, because of their properties, can be reduced to a $2\times 2$ matrix problem in the case of two-mode scattering. In this framework, we provide the distribution of the level spacing of its eigenvalues and show that
this special kind of distribution has no mean (divergent) and is characterized
by a geometrical decay law. We discuss the relation of this distribution to the averaged level spacing
at the edge of the spectrum of $N \times N$ Lorentzian matrices.

\end{abstract}

\maketitle

In random matrix theory, the level spacing, defined as the distance between two adjacent eigenvalues $s=|E_\alpha-E_{\alpha-1}|$, 
 plays an important role in the study of chaotic cavities. Indeed, the distribution of this variable shows whether the system is integrable
(Poissonian statistics for the level distribution $\mathcal{P}(s)$) or non-integrable (strong level repulsion at small $s$)\cite{Mehta}. Moreover, the mean level spacing is an important scaling 
parameter for many physical observalbes. One of the most astonishing success of random matrix theory is the explanation of the distribution 
of the eigen-energies in nuclear spectra. Wigner obtained this distribution from the level spacing of a $2\times 2$ Gaussian matrix. Remarkably, it has been noticed that even though the resulting distribution is 
not exact for the  $N \times N$ case \cite{Mehta}\cite{Forrester}\cite{Gaudin}, it remains an excellent approximation and still referred to as the Wigner surmise distribution.   
In this framework, we will be interested in the distribution of a $2\times 2$ Lorentzian matrix. The choice of this case is very subtle and goes beyond the 
simple mathematical curiosity. The first argument is related to Dyson's circular ensembles (CE). Indeed, in order to obtain a scattering matrix with an ``equal a priori probability ansatz``, we define 
an ensemble of unitary matrices $S$ uniformly distributed according to the Haar measure $P(S)dS= \frac{1}{C}\delta S$. $C$ is a normalization constant.
The scattering matrix $S$ relating the amplitudes of the incoming wave functions to the outgoing ones is related to the scattering Hamiltonian as follows \cite{Weidenmuler1}:
\begin{eqnarray}
 S=1-2\pi i \mathcal{W}^\dagger \frac{1}{E-\mathcal{H}-\Sigma} \mathcal{W}
\end{eqnarray}
The $N \times N$ matrix  $\mathcal{H}$ is the Hamiltonian of the chaotic cavity and $\Sigma$ is the self energy of the channels (the conducting modes). $\mathcal{W}$ is a coupling matrix between the leads (channels) and  
the Hamiltonian of the closed cavity. Thus, for a \textit{finite size} Hamiltonian describing a chaotic cavity, It can be shown \cite{Brouwer} that considering a Lorentzian  distribution for $\mathcal{H}$, with appropriate center and width\cite{Adel}, implies uniformly distributed scattering matrices:
\begin{eqnarray}
 P(\mathcal{H})\propto\frac{1}{\det(\lambda^2+(\mathcal{H}-\mathcal{E})^2)^{\beta(N-1)/2+1}} \Rightarrow S\in \text{CE} \nonumber \\
   \label{DistributionOfH}
\end{eqnarray}
$\beta$ stands for the different symmetry cases ($\beta=1$ in presence of time reversal symmetry TRS, $\beta=2$ without TRS, and $\beta=4$ broken spin-rotational symmetry) 
and N is the size of the Hamiltonian. The center $\mathcal{E}$ and the width $\lambda$ are directly related to the self energy of the channels $\Sigma=\mathcal{E}-i\lambda$.
It is worth stopping at this interesting result and precise that is exact and valid for all the Hamiltonian sizes N. Nevertheless, for very large systems,
it has been shown \cite{Brouwer} that a Gaussian matrix exhibiting a same mean level spacing becomes equivalent to the Lorentzian distribution. Actually, this
result is more general and stands for each distribution $ P(\mathcal{H}) \propto \exp{\{-N tr V(\mathcal{H})\}}$ with a confining potential V generating a smooth mean level spacing 
$\Delta \sim \mathcal{O}(N^{-1})$ when $N\rightarrow\infty$ \cite{Widenmuller}\cite{Beenakker}. It comes out from this remark that the distribution of the local level spacing 
of a $N\times N$ Lorentzian  matrix ($N$ being large) is given by the Wigner surmise when the spacing is taken in the bulk of the spectrum. The $2\times 2$ Lorentzian matrix do not have 
a dense spectrum and therefore do not fulfill this condition. 
Now, we come back to the second argument concerning the choice of a $2\times 2$ matrix. The scattering 
of two independent and equivalent conducting modes is widely considered. In this case, the size of the scattering matrix is $2\times2$. Nevertheless, the Hamiltonian is 
an $N\times N$ matrix where the size $N$ is generally taken very large. Since there is only two modes, we can rewrite the scattering matrix as follows \cite{Adel}\cite{Adel2}:
\begin{eqnarray}
 S=1-2\pi i \tilde{\mathcal{W}}^\dagger \frac{1}{E-\tilde{\mathcal{H}}(E)-\tilde{\Sigma}} \tilde{\mathcal{W}}
\end{eqnarray} 
where $\tilde{\mathcal{H}}(E)$ is an effective Hamiltonian, energy dependent, which has for size $2\times 2$. It is defined as follows $\tilde{\mathcal{H}}(E)=\tau^\dagger\frac{1}{E-\mathcal{H}} \tau$.
$\tilde{\mathcal{W}}$ is a $2\times 2$ sub-matrix of $\mathcal{W}$ containing only the two conducting modes. $\tilde{\Sigma}$ is a $2\times 2$ submatrix of $\Sigma$ containing only the relevant terms.
$\tau$ is $N\times 2$ matrix with $\tau^\dagger_{ij}=\delta_{ij}$. This operation is exact and not an approximation (up to a phase depending on the choice of the scattering potential limits). It allows to reduce  consequently the high  degrees of freedom 
of the original cavity. The amazing thing is that the resulting effective Hamiltonian $\tilde{\mathcal{H}}$ is still Lorentzian. This is due to the following two properties \cite{Brouwer}\cite{Hua}:
\begin{itemize}
 \item If $A$ is Lorentzian with center and width $(\mathcal{E},\lambda)$, $\frac{1}{A}$ is also Lorentzian with $(\frac{\mathcal{E}}{\mathcal{E}^2+\lambda^2},\frac{\lambda}{\mathcal{E}^2+\lambda^2})$. 
 \item If $A$ is Lorentzian, any submatrix of A is Lorentzian with the same width and center. 
\end{itemize}

To summarize, we can say that any $N\times N$ Lorentzian Hamiltonian describing the cavity, boils down to a $2\times 2$ effective Lorentzian matrix problem. This, suggests that the 
transport coefficients (transmission, shot noise $\dots$) are closely related to the statistics of the  two levels of the effective Hamiltonian $\tilde{\mathcal{H}}(E)$.
In what follows, we give the distribution of the level spacing of this $2\times 2 $ Lorentzian matrix.

\subsection{Level spacing distribution}
The starting point to express the level spacing of a $2\times 2$ Lorentzian matrix $\mathcal{H}$ is the  
distribution of the eigenvalues (we call them $x$ and $y$). This is obtained from the distribution Eq. (\ref{DistributionOfH}) by the usual  variable change of the eigenvalues \cite{Brouwer}\cite{Fritz}. We start with the case of
a centered ($\mathcal{E}=0$) and standard ($\lambda=1$) distribution: 
\begin{eqnarray}
P(x,y)=\frac{1}{2\pi} \frac{|x-y|^\beta}{(1+x^2)^{(\beta/2+1)}(1+y^2)^{(\beta/2+1)}}
\end{eqnarray}
$|x-y|^\beta$ is the Jacobian of the variable substitution. It acts as a repulsion term avoiding the eigenvalues to be the same \cite{Mehta}\cite{Fritz}.
The distribution of the level spacing $s$ is therefore expressed as follows:
\begin{eqnarray*}
  \mathcal{P}(s) &=& \iint\delta(s-|x-y|)  \frac{|x-y|^\beta}{[(1+x^2)(1+y^2)]^{(\beta/2+1)}} \frac{dx dy}{2 \pi} \\
\end{eqnarray*}
After a first integration over the delta function, one obtains: 
\begin{eqnarray}
\mathcal{P}(s) =\frac{s^\beta}{2\pi}[p(s)+p(-s)] \label{functionPs}
\end{eqnarray}
where the function  $p(s)$ reads:
\begin{eqnarray}
 p(s)=\int_{-\infty}^{+\infty}   \frac{dx}{[1+x^2]^{(\beta/2+1)} [1+(s-x)^2]^{(\beta/2+1)}}
\end{eqnarray}
It is interesting to notice that $p(s)=(f \ast f)(s)$\\
where $\ast$ is the convolution product and $f$ is the following function: 
\begin{eqnarray} 
 f(x)=\frac{1}{(1+x^2)^{(\beta/2+1)}}. \label{Functionf}
\end{eqnarray}
The convolution product suggests to use the Fourier transform $\mathcal{F}$(noted also $\hat{}  $ ) :
\begin{eqnarray}
\mathcal{F}(p(s))=  \hat{f}^2 \label{1}
\end{eqnarray}
The Fourier transform of $f$ Eq. (\ref{Functionf})  is expressed using the modified Bessel function of the second kind, BesselK \cite{Mathematica}:
\begin{eqnarray}
 \hat{f}(\omega)\propto |\omega|^{\frac{\beta+1}{2}} \text{BesselK}[\frac{\beta+1}{2},|\omega|]
\end{eqnarray}
To obtain the function $p(s)$, we need to do the inverse of the Fourier transform of Eq. (\ref{1}). 

\begin{eqnarray}
p(s)\propto \mathcal{F}^{-1}\left((|\omega|^{\frac{\beta+1}{2}}\text{BesselK}[\frac{\beta+1}{2},|\omega|])^2\right) \label{Eq10}
\end{eqnarray}
The inverse Fourier transform in Eq. (\ref{Eq10}) is expressed using the Gauss hypergeometric function $\mathstrut_2 F_1$ \cite{Mathematica}:

\begin{eqnarray}
p(s)\propto\mathstrut_2 F_1(\frac{3}{2}+\beta,\frac{\beta+2}{2},\frac{\beta+3}{2},-\frac{s^2}{4}) \label{functionFs}
\end{eqnarray}
Now, we are ready to express the probability density function of the level spacing for the three ensembles in a unique compact form deduced directly from Eqs.(\ref{functionPs}) and (\ref{functionFs}) :
\begin{eqnarray} 
\mathcal{P}(s)= C_\beta s^\beta\mathstrut_2 F_1(\frac{3}{2}+\beta,\frac{\beta+2}{2},\frac{\beta+3}{2},-\frac{s^2}{4}) \label{result}
\end{eqnarray}
where the normalization constant $ C_\beta$ is given as follows:
\begin{eqnarray}
C_\beta=\frac{1}{2^\beta \sqrt{\pi}} \frac{\Gamma(\frac{3}{2}+\beta)}{\Gamma(\frac{1+\beta}{2})\Gamma(\frac{3+\beta}{2})}.
\end{eqnarray} 
The behavior of this probability density at small level spacing $(s\rightarrow 0)$ is given by:
\begin{eqnarray}
\mathcal{P}(s) \sim C_\beta s^\beta,
\end{eqnarray}
this is the signature of the level repulsion of each of the three ensembles ($\beta=1,2 \text{ or } 4$) corresponding to different type of symmetry.  This feature is comparable to 
what is found for the Wigner surmise\cite{Mehta}\cite{Fritz}.
The tail of the distribution is more interesting since it is different from all the distributions known so far in a sense it has a geometrical fall off:
\begin{eqnarray}
\mathcal{P}(s)\sim \frac{4}{\pi s^2},
\end{eqnarray}
this tail is $\beta\text{-independent}$ and the exponent  of the fall off law implies the important property of a no mean for the density function.
It means that the mean level spacing defined usually as $\bar{s}=\int s\mathcal{P}(s) ds$ does not exist since the integral diverges. This signifies that despite the distribution is narrow around it's center (for small $\lambda$), 
the eigenvalues can be typically as far as possible.

\section{The distribution of the level spacing at arbitrary width}
It is obvious that changing the center of the distribution does  only shift all the energies by the same amount and therefore the level spacing distribution 
remains unchanged. This is not the case of the  distribution width . Indeed, at a different energy, the width of the Lorentzian changes and 
becomes arbitrary  $\lambda$. We can obtain the distribution at this energy by changing the variables in the integral we started with or just by saying that with
this new width of the distribution, all the eigenvalues are multiplied by a factor $\lambda$ and therefore, the invariant distribution is obtained for the renormalized level spacing  defined as follows:
$$ \mathcal{S}=\frac{ s}{\lambda}$$
Within this definition, the distribution of the normalized variable $\mathcal{S}$ is the same as in Eq. (\ref{result}). 
Usually, we prefer to normalize the lengths in the problem with the mean level spacing $\bar{s}$. Here, since the mean level spacing is divergent, we can not use it for this task.
There is another quantity which can be used as a scaling length. It is also commonly called in literature mean level spacing and noted $\Delta$.
This variable is defined as the inverse of the density of states at the center of the Hamiltonian spectrum:
$\Delta=\frac{1}{\rho(E=\mathbb{\mathcal{E}})}$  
where  the density $\rho$ is defined as $\rho(E)=\sum_{E_i=\{x,y\}} \langle \delta(E-E_i) \rangle $.
This scale $\Delta$ is still defined for the case of a $2\times 2$ matrix and is finite. Nevertheless, it has no anymore the  signification of a mean level spacing 
since the spectrum is not dense. The density of states $\rho$ is $\beta$-independent and reads \cite{Brouwer}:
\begin{eqnarray}
 \rho(E)=\frac{2}{\pi} \frac{\lambda}{(E-\mathcal{E})^2+\lambda^2}
\end{eqnarray}
thus, the scale $\Delta$ at the center of the distribution reads:
\begin{eqnarray}
\Delta=\frac{\pi}{2}\lambda 
\end{eqnarray}
This parameter is equivalent(up to a factor $\pi  \over 2$) to what was stated before about choosing the width $\lambda$ as a simpler and natural scaling parameter. Moreover, by choosing $\lambda$, the factors in
Eq. (\ref{result}) remain unchanged.  
Hereafter, the variable $s$ will  stand for the  level spacing renormalized by the width of the distribution $\lambda$. Within this definition, it is worth keeping in mind the important feature of the level spacing
distribution at small $s$ (resistance to crossing) and at large $s$ (no mean level spacing) which are true for the three ensembles ($\beta=1$, $2$ and $ 4$).
\begin{eqnarray}
\mathcal{P}(s)\sim
    \begin{cases}  C_\beta s^\beta, & \mbox{ if  $s \ll 1$ }\\
                  \frac{4}{\pi s^2 },& \mbox{ if  $ s \gg 1$ }  \label{Dl}
     \end{cases}
\end{eqnarray}
We stress that this geometrical fall-off exhibits a high degree of fluctuation of the level spacing. It is completely different from the usual decay we face in the Wigner surmise  or the semi-poissonian 
ensembles\cite{Pichard}\cite{Bogomolny}.
\subsection{Orthogonal case $\beta=1$}
\begin{figure}
\includegraphics[scale=0.55]{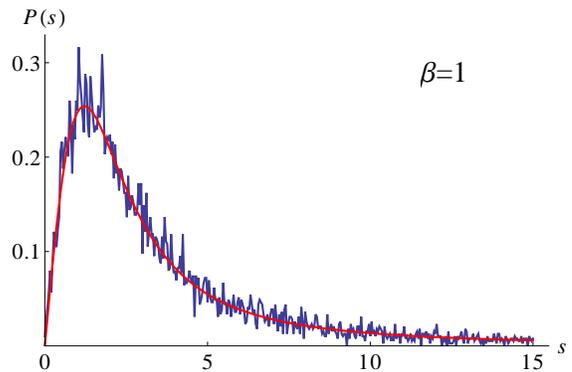} 
\caption{The distribution of the renormalized level spacing of a $2\times 2$ Lorentzian orthogonal matrix ($\beta=1$). The red line is the analytical result given in Eq. (\ref{Orthogonal}). The agreement between theory and numerical simulation is excellent. }
\label{Fig1}
\end{figure}
It may be interesting to simplify formula Eq. (\ref{result}) for each case of the three symmetries ($\beta=1, 2$ and $4$). We start here with the case of time reversal symmetry $\beta=1$. It can be shown that 
the expression of the level spacing distribution boils down to the following form using the complete elliptic integrals of the first and the second kind 
noted respectively  $\mathbb{K}$ and $\mathbb{E}$ \cite{Mathematica}:

\begin{eqnarray}
\mathcal{P}(s)= \left( \frac{8}{\pi} \right)\frac{(-4+s^2)\mathbb{E}(-\frac{s^2}{4})+(4+s^2)\mathbb{K}(-\frac{s^2}{4})}{s(4+s^2)^2} \label{Orthogonal}
\end{eqnarray}
The numerical simulation consisting on sampling $2\times2$ Hamiltonians with a Lorentzian distribution and doing the statistics of the level spacing shows an excellent agreement with formula Eq. (\ref{Orthogonal}) as can be shown in 
Fig. \ref{Fig1}.
The expansion at small and large level spacing can be obtained straightforwardly:

\begin{eqnarray*}
\mathcal{P}(s)\sim
    \begin{cases}
             \frac{3 s}{8}, & \mbox{if $s \ll 1 $ } \\  
             \frac{4}{\pi s^2 }, & \mbox{if $ s\gg 1 $ }
    \end{cases}
\end{eqnarray*}

\section{ Unitary case $\beta=2$}
\begin{figure}
\includegraphics[scale=0.5]{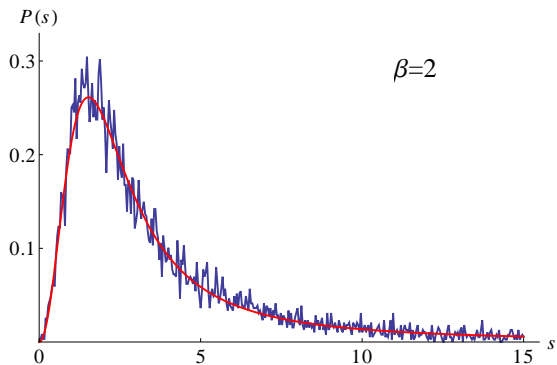}
\caption{The distribution of the renormalized level spacing of a $2\times 2$ Lorentzian Unitary matrix ($\beta=2$). The red line is the analytical result given in Eq. (\ref{UnitaryResult}). The agreement between theory and simulation is excellent.}
\label{Fig2}
\end{figure}
The case of the unitary symmetry ($\beta=2$) can be written in a much simpler formula. Just by taking into account the form of the  Gauss hypergeometric function for even parameter ($\beta$) , one 
finds the following result:
\begin{eqnarray}
 \mathcal{P}(s)=\frac{4}{\pi}\frac{s^2(20+s^2)}{(4+s^2)^3} \label{UnitaryResult}
\end{eqnarray}
The behavior at small and large level spacing is in agreement with formula Eq. (\ref{Dl}):
\begin{eqnarray*}
\mathcal{P}(s)\sim
    \begin{cases}
             \frac{5 s^2}{4\pi}, & \mbox{if $s \ll 1 $ } \\  
             \frac{4}{\pi s^2 }, & \mbox{if $ s\gg 1 $ }
    \end{cases}
\end{eqnarray*}
This distribution law is tested numerically (see Fig. \ref{Fig2}).
Again, we notice that this distribution has no mean level spacing.\\
\subsection{Symplectic ensemble $\beta=4$}
\begin{figure}
\includegraphics[scale=0.55]{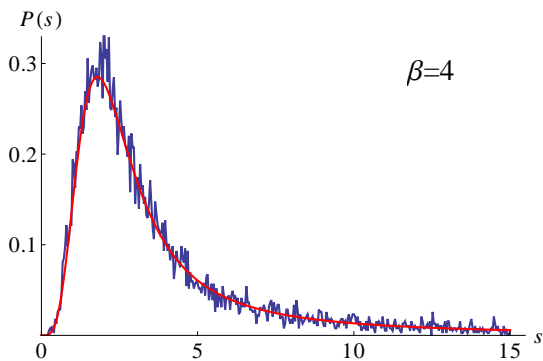}
\caption{The distribution of the renormalized level spacing of a $4\times 4$ Lorentzian matrix with the symplectic symmetry ($\beta=4$). (notice the factor $2$ due to the degeneracy of the eigenvalues). The red line is the analytical result given by Eq. (\ref{DisSymplectic}). The agreement between theory and simulation is excellent.}
\label{FigSymplectic}
\end{figure}
Again, the distribution of the level spacing for the symplectic ensemble is much simpler and can be written in a fractional form:

\begin{eqnarray}
\mathcal{P}(s)=\frac{4 s^4}{\pi}\frac{(336+24 s^2+ s^4)}{(4+s^2)^5} \label{DisSymplectic} 
\end{eqnarray}
The expansion at small and large level spacing is straightforward:
$$
\mathcal{P}(s) \sim \begin{cases} \frac{21 s^4}{16\pi}, & \mbox{if  $ s \ll 1$ } \\ 
               \frac{4}{\pi s^2 }, & \mbox{if  $s\gg 1$} \end{cases}
$$
Now, it becomes clear that the tail of the level spacing distribution is $\beta$-independent.\\
The numerical simulation showing an excellent agreement with the distribution Eq. (\ref{DisSymplectic}) is given in Fig. \ref{FigSymplectic}. 
\subsection{$N \times N$ Lorentzian Hamiltonian}
\begin{figure}
\includegraphics[scale=0.5]{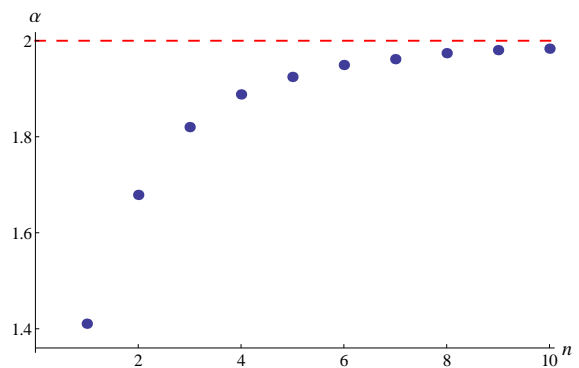}
\caption{The exponent of the geometrical fall-off  $1/x^\alpha$ of an  $N\times N$ Lorentzian matrix as a function of the number of levels $n\equiv N_e$ taken to define the averaged mean level spacing. The levels are  
taken at the edge of the spectrum starting from the largest eigenvalue.  }
\label{Fig4}
\end{figure}

\begin{figure}
\includegraphics[scale=0.5]{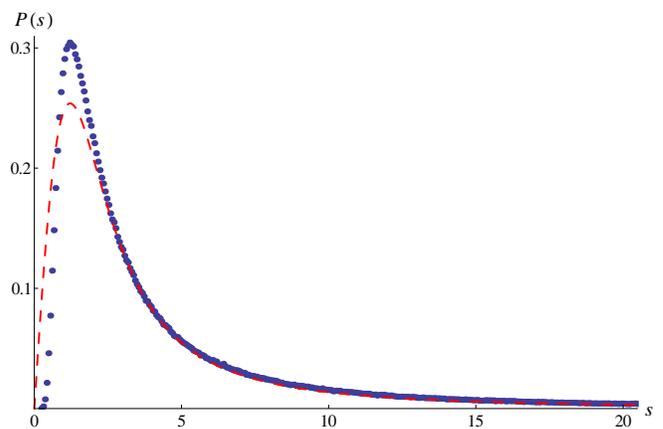}
\caption{Distribution of the averaged level spacing at the edge of the spectrum for $N_e=8$ and $\beta=1$ (Blue curve). The size of the matrix is $N=40$. The red curve represents the theoretical 
result for $2\times2$ orthogonal Lorentzian ensembles. The figure shows a very good agreement 
in the description of the tail which suggests a very high degree of fluctuation and the absence of the mean for this distribution.}
\label{EdgeOf_spectrum}
\end{figure}

\begin{figure}
\includegraphics[scale=0.5]{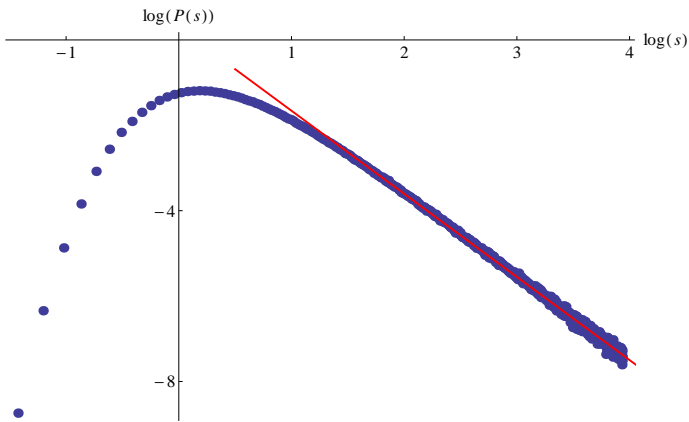}
\caption{$log(\mathcal{P}(s))$ as a function of $log(s)$. The linear decay for large values of $s$ is characteristic of a geometrical fall-off of the distribution $\mathcal{P}(s)$. Red curve is the linear fit of the numerical data.
It has for equation: $y=-1.95x+0.30$. The absolute value of the slope $\alpha=1.95$ represents the exponent of the geometrical fall-off.}
\label{LogLog}
\end{figure}

It is well known \cite{Brouwer} that a large matrix taken from the Lorentzian ensemble is equivalent to a Gaussian matrix sharing the same mean level spacing $\Delta$ at the bulk of the spectrum. This result 
was obtained by comparing the cluster functions \cite{Brouwer} and one can understand this as a special case of the general idea of the bulk spectrum universality \cite{Widenmuller}. Therefore, it is easy to 
test that the level spacing taken from the bulk spectrum is well fitted by the Wigner surmise. The situation is not the same at the edge (tail) of the spectrum: the fluctuations are higher and 
the mean level spacing diverges. Fig. \ref{EdgeOf_spectrum} shows the distribution of the averaged level spacing of  an Orthogonal Lorentzian matrix at the edge of the spectrum, defined as :
$$ S=\frac{1}{N_e-1}\sum^{N_e}_{i=1} |E_i-E_{i-1}|$$ 
where the sum is taken over some  $N_e$ levels at the edge of the spectrum (the levels $E_i$ are ordered such that $E_1$ represents the largest eigenvalue.). It shows clearly that the tail falls off  with a geometrical 
law comparable to that of the $2\times2$ matrix studied in the previous sections. This can be more visible in the plot of $\log(\mathcal{P})$ as a function of $\log(s)$ shown in Fig. \ref{LogLog}. The decay law  seems quite well fitted 
by $\frac{1}{s^\alpha}$ with $\alpha$ a parameter depending on the number of levels considered in the edge. As shown in Fig. \ref{EdgeOf_spectrum}, the law of the averaged level spacing has no mean 
since mostly we have $\alpha <2$. It is now important to define  where the edge of the spectrum roughly starts. This can be seen as the levels where the distribution of the spacing between two consecutive 
levels stops to be well fitted by a Wigner surmise. In Fig. \ref{Fig4}, we give the exponent $\alpha$ for different numbers $N_e$ of levels taken  from the edge of the spectrum. We see that the exponent $\alpha$ approaches the value $\alpha=2$ when we 
increase $N_e$. We stress that increasing $N_e$ needs larger values of $N$ in order to count only the eigenvalues at the edge.  \\
This behavior at the edge of the spectrum suggests a new physics different from that of the bulk. This was already noticed in \cite{Adel}\cite{Adel2} where the statistics of thermopower and the delay time 
of a chaotic cavity were found to be different in the two situations corresponding to a Fermi energy lying in the bulk of the spectrum or in its edge. Moreover, some features and distributions can be directly obtained from the 
$2\times 2$ Lorentzian Hamiltonian instead of the original $N\times N$ matrix. Indeed, the distribution of the Seebeck coefficient \cite{Adel} and the Wigner's time \cite{Adel2} at the edge 
of the spectrum were found directly by considering the $2\times2$ Lorentzian Hamiltonian.     
 
\subsection{Conclusion:}
We gave the exact form of the level spacing  distribution of a $2\times2$ Lorentzian matrix.
This kind of matrices appear in chaotic scattering problems as an effective matrix replacing large matrices describing chaotic cavities.
The tail of the distribution is slowly decaying and leads to the absence of the mean level spacing.

\section{**********}
The author is grateful to J. L. Pichard and K. Muttalib for introducing him to this subject and to RMT in general. He would like to thank G. Fleury for valuable discussions and remarks.
 \\
The author acknowledges partial support of the R\'egion Basse Normandie.

\end{document}